\documentclass[12pt,thmsa]{article}
\usepackage{sw20lart}

\input{tcilatex}
\begin{document}

\begin{center}
{\Large A convenient method to prepare the effective pure state in a quantum
ensemble }
\end{center}

\ 

\begin{center}
Xijia Miao\textbf{\ }

Laboratory of Magnetic Resonance and Atomic and Molecular Physics, Wuhan
Institute of Physics and Mathematics, The Chinese Academy of Sciences, Wuhan
430071, People$^{^{\prime }}$s Republic of China; Fax: 86 27 87885291,
E-mail: miao@nmr.whcnc.ac.cn

\ 

\textbf{Abstract}
\end{center}

A simple method is proposed to prepare conveniently the effective pure state 
$|00...0\rangle \langle 0...00|$ with any number of qubits in a quantum
ensemble. The preparation is based on the temporal averaging (Knill, Chuang,
and Laflamme, Phys. Rev. A 57, 3348 (1998)). The quantum circuit to prepare
the effective pure state is designed in a unified and systematical form and
is explicitly decomposed completely into a product of a series of one-qubit
quantum gates and the two-qubit diagonal quantum gates. The preparation
could be programmed and implemented conveniently on an NMR quantum computer. 
\newline
\newline
PASC number(s): 03.67.Lx, 76.60.-k\newline
\newline
\textbf{1. Introduction }

Quantum computers can solve certain problems that can not be solved by any
classical digital computers [1-5]. Particularly, the prime factorization of
a large number can be performed in a polynomial time on a quantum computer
[3]. This has stimulated the experimental work of quantum computation in
various physical disciplines including quantum optics [6], trapped ions [7],
nuclear spin system [8], superconducting Josephson junctions [9], and so
forth. Quantum computation is usually performed in a pure quantum state at
any time [10]. However, recently it has been shown that quantum computation
may be implemented in a spin quanutm ensemble at a finite temperature by
using nuclear magnetic resonance (NMR) techniques [11, 12]. After that,
there has been a flood of experimental work of ensemble quantum computation
on few-qubit systems [13-20]. The key of the approach is that a quantum
ensemble can be prepared in certain mixed states, i.e., the effective pure
states [11] or the pseudopure states [12]. An effective pure state is a
mixed state in a quantum ensemble that behaves for all computational
purposes as a pure state, that is, the effective pure state is isomorphic to
the pure state. Therefore, the reversible quantum computation can be
performed in an effective pure state just like in a pure quantum state. The
key step in the approach is then the preparation of the effective pure
states. There are several methods to prepare the effective pure states or
the pseudopure states. These include logical labeling [11, 21], spatial
averaging [12, 22], temporal averaging [23], spectral labeling [17], and so
forth. The temporal averaging method is really a technique much similar to
the time averaging, coherent accumulation, and phase cycling techniques in
NMR experiments [24, 25], but used for preparation of the effective pure
states. This method has been investigated thoroughly by Knill, Chuang, and
Laflamme [23]. The method such as the exhaustive averaging method usually
involves cyclicly permuting the nonground states in $2^{n}-1$ different ways
such that the average of the prepared states is an effective pure state.
Therefore, number of experiments to prepare the effective pure state grows
exponentially as the qubit numbers $n$, showing that this method may be
reasonable to consider implementing the preparation for small numbers of
qubits. The advantage of the method over other preparations [11, 12] may be
that it is a simple method to give a higher signal-to-noise ratio for the
NMR experimental measurement.

In this paper a simple method is proposed to prepare conveniently the
effective pure state $\rho _{eff}=\lambda |00...0\rangle \langle 0...00|$ in
a spin quantum ensemble that consists of a large amount of two-state n-spin
(I=1/2) quantum systems corresponding to the pure quantum ground state $%
|00...0\rangle $ of the quantum system. It is also based on the temporal
averaging [23]. However, the preparation starts generally from the specific
reduced density operator of the longitudinal n-spin order component $\sigma
(0)=2^{n-1}I_{1z}I_{2z}...I_{nz}$ [26, 27] that can be easily obtained from
the equilibrium magnetization in the spin quantum ensemble (called also the
spin system without confusion). The quantum circuit to prepare the effective
pure state with any number of qubits is designed in a unified and
systematical form. It is then explicitly decomposed completely into a
product of a series of one-qubit quantum gates and the two-qubit diagonal
gates [26]. Therefore, the preparation could be programmed and implemented
conveniently on an NMR quantum computer. \newline
\newline
\textbf{2. Preparation of the effective pure states}

An effective pure state or pseudopure state $\rho _{eff}(\Psi )$ in a
quantum ensemble corresponding to the pure quantum state $|\Psi \rangle $
may be defined as [11, 12]

$\qquad \qquad \qquad \rho _{eff}(\Psi )=(1-\lambda )E+\lambda |\Psi \rangle
\langle \Psi |\qquad \qquad \qquad \qquad \qquad \quad \ (1)$ \newline
where the normalization factor is omitted without confusion, $\lambda $ is a
real constant, and $E$ is $N$ $(=2^{n})$-dimensional unity operator. Because
the unity operator is unobservable and keeps unchanged when it is acted on
by an arbitrary unitary transformation the mixed state $\lambda |\Psi
\rangle \langle \Psi |$ and the effective pure state $\rho _{eff}(\Psi )$
cannot be distinguished by standard NMR experiments. Therefore, the process
of NMR ensemble quantum computing can be really characterized completely
through the reduced effective pure state $\sigma _{eff}=\lambda |\Psi
\rangle \langle \Psi |.$ In order to make use of the massive quantum
parallelism [2] the superposition in a quantum system is usually created at
the first step in quantum computing, while the superposition in a spin
quantum ensemble can be created conveniently by acting the Walsh-Hadamard
transform $W$ [5] on the effective pure state corresponding to the pure
quantum groundstate $|00...0\rangle $ in the quantum system:

$\qquad \qquad \qquad \qquad \sigma _{eff}=\lambda |00...0\rangle \langle
0...00|\qquad \qquad \qquad \qquad \qquad \qquad \ \ \ (2)$ \newline
Thus, in the paper a simple method is proposed how to prepare conveniently
the effective pure state of Eq.(2) in a spin quantum ensemble with any
number of qubits.

First of all, the starting density operator $\rho (0)$ of the spin quantum
ensemble is subjected to a unitary transformation whose unitary operator $%
\exp (-i\alpha Q)$ is constructed with the Hermitian operator $Q$ defined by 
$Q_{ij}=1$ for all indexes $i$ and $j:$

$\exp (-i\alpha Q)\rho (0)\exp (i\alpha Q)=\rho (0)-\dfrac{1-\cos (\alpha N)%
}{N}[\rho (0),Q]_{+}$

$\qquad +i\dfrac{\sin (\alpha N)}{N}[\rho (0),Q]_{-}+\dfrac{[1-\cos (\alpha
N)]^{2}+\sin ^{2}(\alpha N)}{N^{2}}Q\rho (0)Q\qquad \quad \ \ (3)$ \newline
where $[\rho (0),Q]_{\pm }=$ $\rho (0)Q\pm Q\rho (0)$ and the following
expansion for the unitary operator exp($\pm i\alpha Q)$ has been introduced
in the above transformation,

\qquad \qquad \qquad exp($\pm i\alpha Q)=E-\dfrac{1-\exp (\pm i\alpha N)}{N}%
Q\qquad \qquad \qquad \qquad \quad \ (4)$ \newline
It follows from the definition of the operator $Q$, that is, $Q_{ij}=1$ for
all indexes $i$ and $j$, that the product $Q\rho (0)Q$ can be reduced to the
simple form

$\qquad \qquad \qquad \qquad \qquad Q\rho (0)Q=Qf\{\rho (0)\}\qquad \qquad
\qquad \qquad \qquad \quad \ \ (5)$ \newline
where the function $f\{\rho (0)\}$ is defined as

$\qquad \qquad \qquad \qquad \qquad f\{\rho (0)\}=\stackrel{N}{\stackunder{%
k,l=1}{\sum }}\rho _{kl}(0)\qquad \qquad \qquad \qquad \qquad \quad \ (6)$ 
\newline
For convenience, one writes conveniently the density operator as the general
form

$\qquad \qquad \qquad \qquad \qquad \rho (0)=Tr\{E\}^{-1}E+\sigma (0),$ $%
\qquad \qquad \qquad \qquad \quad (7)$\newline
where $\sigma (0)$ is the traceless reduced density operator, $Tr\{\sigma
(0)\}=0$. For an NMR spin quantum\ ensemble the high-temperature
approximation holds and equation (7) is met. Obviously, it proves easily
with the aid of Eq.(7) that equation (3) still holds when replacing the
reduced density operator $\rho (0)$ with the density operator $\sigma (0)$.
To simplify further Eq.(3) the operator $Q$ is decomposed explicitly as

$\qquad \qquad \qquad \qquad Q=N\exp (-i\frac{\pi }{2}F_{y})D_{0}\exp (i%
\frac{\pi }{2}F_{y})\qquad \qquad \qquad \quad (8)$ \newline
where the operator $F_{\mu }=\stackrel{n}{\stackunder{k=1}{\sum }}I_{k\mu },$
$(\mu =x,y,z),$ $I_{k\mu }=\frac{1}{2}\sigma _{k\mu },$ $\sigma $ is the
Pauli's operator and the diagonal operator $D_{s}$ is defined as

$D_{s}=Diag[0,...,0,1,0,...,0],$ $([D_{s}]_{ss}=1;$ $[D_{s}]_{ii}=0,$ $i\neq
s).$\newline
Then by exploiting Eqs.(5) and (8) one obtains from Eq.(3) that

$\dfrac{[1-\cos (\alpha N)]^{2}+\sin ^{2}(\alpha N)}{N}f\{\rho (0)\}D_{0}$

$=\exp (-iN\alpha D_{0})\rho _{+}(0)\exp (iN\alpha D_{0})-\rho _{+}(0)$

$\quad +(1-\cos (\alpha N))[\rho _{+}(0),D_{0}]_{+}-i\sin (\alpha N))[\rho
_{+}(0),D_{0}]_{-}\qquad \qquad (9)$\newline
where $\rho _{+}(0)=$ $\exp (i\frac{\pi }{2}F_{y})\rho (0)\exp (-i\frac{\pi 
}{2}F_{y}).$ In particular, if the parameter $\alpha $ in Eq.(3) is chosen
suitably so that $\cos (\alpha N)=-1$ and $\sin (\alpha N)=0,$ then equation
(9) can be further written as with the help of Eq.(7)

$\dfrac{4}{N}f\{\sigma (0)\}D_{0}=\exp (-iN\alpha D_{0})\sigma _{+}(0)\exp
(iN\alpha D_{0})$

$\qquad \qquad \qquad \qquad -\sigma _{+}(0)+2[\sigma
_{+}(0),D_{0}]_{+}\qquad \qquad \qquad \qquad \qquad \ \ (10)$ \newline
where $\sigma _{+}(0)=$ $\exp (i\frac{\pi }{2}F_{y})\sigma (0)\exp (-i\frac{%
\pi }{2}F_{y}).$ Now, the function $f\{\sigma (0)\}=0$ if the density
operator $\sigma (0)$ is taken as an arbitrary operator of the operator set $%
G=\{I_{ky},$ $I_{kz},$ $2I_{ky}I_{lx,}$ $2I_{ky}I_{ly},$ $2I_{ky}I_{lz,}$ $%
2I_{kz}I_{lx},...\},$ which does not contain all the x-components and unity
operator of the operator set $G_{x}=\{E,$ $I_{kx},$ $2I_{kx}I_{lx},$ $%
4I_{kx}I_{lx}I_{mx},...\}.$ Then in this case equation (10) can be further
simplified to the form

$\qquad \qquad \exp (-iN\alpha D_{0})\sigma _{+}(0)\exp (iN\alpha
D_{0})-\sigma _{+}(0)$

$\qquad \qquad \qquad =-2[\sigma _{+}(0),D_{0}]_{+}\qquad \qquad \qquad
\qquad \qquad \qquad \qquad \quad (11)$ \newline
When the starting density operator $\sigma (0)$ is taken as an arbitrary
operator of set $G_{x}$ equation (10) becomes an identity although in this
case $f\{\sigma (0)\}\neq 0$. According to the definition of the diagonal
operator $D_{0}$ and $\cos (\alpha N)=-1,$ $\sin (\alpha N)=0$ it turns out
easily that

$\qquad \qquad \exp (\pm iN\alpha D_{0})=Diag[-1,1,...,1]=-R$ \qquad \qquad
\qquad \ \ $(12)\newline
$where $R$ is the phase-shift operation defined in the Grover quantum search
algorithm [5]. By exploiting the definition of diffusion transform $D=WRW$
in the Grover quantum search algorithm [5], equation (11) is rewritten as

$\qquad (2/N)[\sigma _{+}(0),ND_{0}]_{+}=\sigma _{+}(0)-WDW\sigma
_{+}(0)WDW\qquad \qquad \ \ (13)$ \newline
Equation (13) is the key to prepare conveniently the effective pure state $%
\rho _{eff}=\lambda |00...0\rangle \langle 00...0|\ $in the paper.

It follows from the definition of the operator $D_{0}$ that the diagonal
operator $D_{0}$ is really an effective pure state $\rho
_{eff}=|00...0\rangle \langle 0...00|=$ $D_{0}.$ Because $D_{0}$ is a
diagonal operator it can be expanded as a sum of the base operators of the
longitudinal magnetization and spin order ($LOMSO$) operator subspace of the
Liouville operator space of the n-spin (I=1/2) system [17, 26, 27]

$ND_{0}=E+\stackrel{n}{\stackunder{k=1}{\sum }}2I_{kz}+\stackrel{n}{%
\stackunder{l>k=1}{\sum }}4I_{kz}I_{lz}+\stackrel{n}{\stackunder{m>l>k=1}{%
\sum }}8I_{kz}I_{lz}I_{mz}+...\qquad \ (14a)$

$\qquad \quad \ \ =(E_{1}+2I_{1z})\bigotimes (E_{2}+2I_{2z})\bigotimes
...\bigotimes (E_{n}+2I_{nz})\qquad \quad \qquad \ \ (14b)$ \newline
where $E_{k}$ is $2\times 2-$dimensional unity operator of the $k$th qubit
in the spin system. In order to extract the effective pure state from the
term $[\sigma _{+}(0),ND_{0}]_{+}$ on the left-hand side of Eq.(13) the
density operator $\sigma _{+}(0)$ needs to be chosen suitably. First, a
series of the density operators $\sigma _{+}(0)$ are suitably chosen, where
each density operator $\sigma (0)$ is only taken as an arbitrary operator of
set G so that $f\{\sigma (0)\}=0$ can be met, then the term $[\sigma
_{+}(0),ND_{0}]_{+}$ is expanded explicitly by inserting each $\sigma
_{+}(0) $ into the term. By exploiting these expansions one can construct
explicitly the effective pure state $\rho _{eff}=\lambda |00...0\rangle
\langle 00...0|. $ The detailed procedure to prepare conveniently and
explicitly the effective pure state based on equation (13) is illuminated
below. \newline
\newline
\textbf{2.1 The logical labeling effective pure state}

It has been shown that the following state, called the logical labeling
effective pure state, can be thought of as an effective pure state if one of
the qubits in the quantum system is used as a logical label [23]

$\qquad \qquad \rho _{lb}=$ $\delta E+\lambda (|00...0\rangle \langle
0...00|-|11...1\rangle \langle 1...11|).\qquad \qquad \qquad (15)$ \newline
It will be shown below that the logical labeling effective pure state of
Eq.(15) in the n-spin system can be prepared on the basis of Eq.(13). The
density operator $\sigma _{+}(0)$ in Eq.(13) is chosen simply as

$\qquad \qquad \qquad \qquad \ \sigma _{+}(0)=2^{n-1}I_{1y}I_{2y}...I_{ny}$
\qquad $\qquad \qquad \qquad \quad \qquad (16)$\newline
Obviously, the function $f\{\sigma (0)\}=0.$ It is easy to turn out by using
Eq.(14) that

$\ \ [\sigma
_{+}(0),ND_{0}]_{+}=2^{n-1}i^{n}I_{1}^{-}I_{2}^{-}...I_{n}^{-}+2^{n-1}(-i)^{n}I_{1}^{+}I_{2}^{+}...I_{n}^{+}\qquad \qquad (17) 
$

$
\begin{array}{l}
=2^{n-1}(I_{1}^{-}I_{2}^{-}...I_{n}^{-}+I_{1}^{+}I_{2}^{+}...I_{n}^{+}), \\ 
=2^{n-1}i(I_{1}^{-}I_{2}^{-}...I_{n}^{-}-I_{1}^{+}I_{2}^{+}...I_{n}^{+}), \\ 
=-2^{n-1}(I_{1}^{-}I_{2}^{-}...I_{n}^{-}+I_{1}^{+}I_{2}^{+}...I_{n}^{+}), \\ 
=-2^{n-1}i(I_{1}^{-}I_{2}^{-}...I_{n}^{-}-I_{1}^{+}I_{2}^{+}...I_{n}^{+}),
\end{array}
$ $
\begin{array}{l}
\text{if }n=4m \\ 
\text{if }n=4m+1 \\ 
\text{if }n=4m+2\text{ } \\ 
\text{if }n=4m+3.
\end{array}
$ \newline
The right-hand side of Eq.(17) is actually an n-qubit maximum entanglement
state of the system. Therefore, one can prepare conveniently the maximum
entanglement state through Eq.(13). Now, if $n$ is an even number one makes
a unitary transformation on the term $[\sigma _{+}(0),ND_{0}]_{+}$ to
convert it into the form corresponding to $n$ being an odd number

$\exp (-i\theta
F_{z})(I_{1}^{-}I_{2}^{-}...I_{n}^{-}+I_{1}^{+}I_{2}^{+}...I_{n}^{+})\exp
(i\theta F_{z})=$

$\qquad \qquad \qquad
i(I_{1}^{-}I_{2}^{-}...I_{n}^{-}-I_{1}^{+}I_{2}^{+}...I_{n}^{+})$ \qquad
\qquad \qquad $\newline
$where $n\theta =\pi /2$, that is, $\exp (\pm in\theta )=\pm i.$ Therefore,
no matter that $n$ is either an odd number or an even number one can prepare
further the effective pure state by starting from the $n$-order
multiple-quantum coherence $%
2^{n-1}i(I_{1}^{-}I_{2}^{-}...I_{n}^{-}-I_{1}^{+}I_{2}^{+}...I_{n}^{+}).$ By
making a unitary transformation on the multiple-quantum coherence one
obtains the logical labeling effective pure state of Eq.(15) without the
unity operator term

$\exp (-i\frac{\pi }{4}\times
2^{n}I_{1x}I_{2x}...I_{nx})2^{n-1}i(I_{1}^{-}I_{2}^{-}...I_{n}^{-}-I_{1}^{+}I_{2}^{+}...I_{n}^{+}) 
$

$\times \exp (i\frac{\pi }{4}\times 2^{n}I_{1x}I_{2x}...I_{nx})$

$=\frac{1}{2}N(|00...0\rangle \langle 0...00|-|11...1\rangle \langle
1...11|)\qquad \qquad \qquad \qquad \qquad \qquad \quad \ (18)$ \newline
On the other hand, by using Eq.(13) one can design the quantum circuit to
prepare conveniently the logical labeling effective pure state of Eq.(18).
As an example, consider the case of $n=4m+1,$ by inserting Eq.(16) into
Eq.(13) one obtains the logical labelng effective pure state of Eq.(18)

$\rho _{lb}=$ $(|00...0\rangle \langle 0...00|-|11...1\rangle \langle
1...11|)$

$=2^{n-1}I_{1z}I_{2z}...I_{nz}-\exp (-i\frac{\pi }{4}\times
2^{n}I_{1x}I_{2x}...I_{nx})WDW\exp (i\frac{\pi }{2}F_{x})$

$\times 2^{n-1}I_{1z}I_{2z}...I_{nz}\exp (-i\frac{\pi }{2}F_{x})WDW\exp (i%
\frac{\pi }{4}\times 2^{n}I_{1x}I_{2x}...I_{nx})\qquad \qquad \ (19)$ 
\newline
Experimentally, starting from the reduced density operator, i..e., the
longitudinal n-spin order component $\sigma (0)=2^{n-1}I_{1z}I_{2z}...I_{nz}$
that can be prepared from the thermal equilibrium state in an n-spin quantum
ensemble one performs two different experiments to prepare the logical
labelng effective pure state $\rho _{lb}$ of Eq.(18) according to Eq.(19),
one experiment with the identical operation $E$ and another with the unitary
operation $\exp (-i\frac{\pi }{4}\times 2^{n}I_{1x}I_{2x}...I_{nx})WDW\exp (i%
\frac{\pi }{2}F_{x})$. In comparison with the flip and swap method [23] to
prepare the logical labeling effective pure state of Eq.(15) the present
method is still available even when the number of qubits $(n)$ and the
polarization $(\delta )$ of the quantum system satisfy $n\delta \sim 1$ or
when the initial state of the system does not have approximate inversion
symmetry, but as shown in Ref.[23], in these cases the flip and swap method
is failure. \newline
\newline
\textbf{2.2 The effective pure states}\newline
\textbf{(a) A two-spin system}

The density operator $\sigma _{+}(0)$ is chosen as $\sigma
_{+}(0)=2I_{1y}I_{2z},$ $2I_{1z}I_{2y},$ respectively. One obtains with the
help of Eq.(14)

$\exp (-i\frac{\pi }{2}I_{1x})[2I_{1y}I_{2z},ND_{0}]_{+}\exp (i\frac{\pi }{2}%
I_{1x})$

$=\frac{1}{2}ND_{0}-\frac{1}{2}(E_{1}-2I_{1z})\bigotimes (E_{2}+2I_{2z}),$%
\qquad $\qquad \qquad \qquad \qquad \qquad (20a)$

$\exp (-i\frac{\pi }{2}I_{2x})[2I_{1z}I_{2y},ND_{0}]_{+}\exp (i\frac{\pi }{2}%
I_{2x})$

$=\frac{1}{2}ND_{0}-\frac{1}{2}(E_{1}+2I_{1z})\bigotimes (E_{2}-2I_{2z})$
\qquad $\qquad \qquad \qquad \qquad \qquad (20b)$ \newline
By plus Eqs.(20a) and (20b) and then exploiting Eq.(13) one gets the
effective pure state of the two-spin system as follows

$\rho _{eff}=2D_{0}=\frac{1}{2}E+2I_{1z}I_{2z}$

$-\exp (-i\frac{\pi }{2}I_{1x})WDW\exp (i\frac{\pi }{2}%
I_{1x})(2I_{1z}I_{2z})\exp (-i\frac{\pi }{2}I_{1x})WDW\exp (i\frac{\pi }{2}%
I_{1x})$

$-\exp (-i\frac{\pi }{2}I_{2x})WDW\exp (i\frac{\pi }{2}%
I_{2x})(2I_{1z}I_{2z})\exp (-i\frac{\pi }{2}I_{2x})WDW\exp (i\frac{\pi }{2}%
I_{2x})$ \newline
In NMR experiments of ensemble quantum computation the starting reduced
density operator $\sigma (0)=2I_{1z}I_{2z}$ is first prepared from the
equilibrium state in the coupled two-spin (I=1/2) system. By starting from
the density operator $\sigma (0)=2I_{1z}I_{2z}$, one performs three
different experiments to create three different mixed states, one experiment
with identity transformation $E$ and the other two with the unitary
transformations: $\exp (-i\frac{\pi }{2}I_{1x})WDW\exp (i\frac{\pi }{2}%
I_{1x})$ and $\exp (-i\frac{\pi }{2}I_{2x})WDW\exp (i\frac{\pi }{2}I_{2x})$,
respectively. Then the created state $\sigma (0)$ $=$ $2I_{1z}I_{2z}$ minus
the other two created states will give the effective pure state $2D_{0}-%
\frac{1}{2}E$. \newline
\newline
\textbf{(b) A three-spin system}

The density operator $\sigma _{+}(0)$ is chosen as $\sigma
_{+}(0)=4I_{1y}I_{2z}I_{3z}$, $4I_{1z}I_{2y}I_{3z}$, $4I_{1z}I_{2z}I_{3y},$
and $4I_{1y}I_{2y}I_{3y},$ respectively. With the help of Eq.(14) it easily
turns out for the density operator $\sigma _{+}(0)=4I_{ky}I_{lz}I_{mz}$ $($%
cyclicly permuting $k,l,m)$ that

$\frac{2}{N}\exp (-i\frac{\pi }{2}I_{kx})[4I_{ky}I_{lz}I_{mz}$, $%
ND_{0}]_{+}\exp (i\frac{\pi }{2}I_{kx})$

$=\frac{2}{N}\{\frac{1}{2}(E_{k}+2I_{kz})\bigotimes
(E_{l}+2I_{lz})\bigotimes (E_{m}+2I_{mz})$

$\quad -\frac{1}{2}(E_{k}-2I_{kz})\bigotimes (E_{l}+2I_{lz})\bigotimes
(E_{m}+2I_{mz})\}$

$=4I_{kz}I_{lz}I_{mz}-\exp (-i\frac{\pi }{2}I_{kx})WDW\exp (i\frac{\pi }{2}%
I_{kx})$

$\qquad \times (4I_{kz}I_{lz}I_{mz})\exp (-i\frac{\pi }{2}I_{kx})WDW\exp (i%
\frac{\pi }{2}I_{kx})$ \qquad $\qquad \qquad \qquad (21a)$\newline
and for the density operator $\sigma _{+}(0)=4I_{ky}I_{ly}I_{my}$ that

$\frac{2}{N}\exp (-i\frac{\pi }{4}\times
8I_{kx}I_{lx}I_{mx})[4I_{ky}I_{ly}I_{my}$, $ND_{0}]_{+}\exp (i\frac{\pi }{4}%
\times 8I_{kx}I_{lx}I_{mx})$

$=\frac{2}{N}\{-\frac{1}{2}(E_{k}+2I_{kz})\bigotimes
(E_{l}+2I_{lz})\bigotimes (E_{m}+2I_{mz})$

$\quad +\frac{1}{2}(E_{k}-2I_{kz})\bigotimes (E_{l}-2I_{lz})\bigotimes
(E_{m}-2I_{mz})\}$

$=-4I_{kz}I_{lz}I_{mz}$

$-\exp (-i\frac{\pi }{4}\times 8I_{kx}I_{lx}I_{mx})WDW\exp (i\frac{\pi }{2}%
F_{x})$

$\times (4I_{kz}I_{lz}I_{mz})\exp (-i\frac{\pi }{2}F_{x})WDW\exp (i\frac{\pi 
}{4}\times 8I_{kx}I_{lx}I_{mx})$ \qquad \qquad \qquad $(21b)$\newline
where $F_{x}=I_{kx}+I_{lx}+I_{mx}$ and $N=2^{3}.$ Then exploiting Eqs.(21)
the effective pure state in the three-spin system can be built up as follows

$\rho _{eff}=2^{2}D_{0}=\frac{1}{2}E+(2^{2}-1)2^{2}I_{1z}I_{2z}I_{3z}$

$-\stackrel{3}{\stackunder{k=1}{\sum }}\{\exp (-i\frac{\pi }{2}%
I_{kx})WDW\exp (i\frac{\pi }{2}I_{kx})$

$\qquad \times (2^{2}I_{1z}I_{2z}I_{3z})\exp (-i\frac{\pi }{2}I_{kx})WDW\exp
(i\frac{\pi }{2}I_{kx})\}$

$-$ $\exp (-i\frac{\pi }{4}\times 8I_{1x}I_{2x}I_{3x})WDW\exp (-i\frac{\pi }{%
2}F_{x})$

$\qquad \times (2^{2}I_{1z}I_{2z}I_{3z})\exp (i\frac{\pi }{2}F_{x})WDW\exp (i%
\frac{\pi }{4}\times 8I_{1x}I_{2x}I_{3x}).$ \newline
Experimently, one first prepares the starting reduced density operator $%
\sigma (0)=4I_{1z}I_{2z}I_{3z}$ from the equilibrium magnetization in a
coupled three-spin system. Then only five different experiments are
performed in order to prepare the effective pure state $(2^{2}D_{0}-\frac{1}{%
2}E).$ If the experimental signal-to-noise ratio is high enough in the spin
system the contribution of component $(2^{2}-1)2^{2}I_{1z}I_{2z}I_{3z}$ to
the effective pure state can be obtained directly by amplifying the
amplitude of the longitudinal three-spin order component $%
2^{2}I_{1z}I_{2z}I_{3z}$. \newline
\newline
\textbf{(c) A four-spin system}

The density operator $\sigma _{+}(0)$ is chosen as $\sigma
_{+}(0)=8I_{1y}I_{2z}I_{3z}I_{4z}$, \newline
$8I_{1z}I_{2y}I_{3z}I_{4z}$, $8I_{1z}I_{2z}I_{3y}I_{4z},$ $%
8I_{1z}I_{2z}I_{3z}I_{4y},$ and $8I_{1z}I_{2y}I_{3y}I_{4y}$, $%
8I_{1y}I_{2z}I_{3y}I_{4y},$\newline
$8I_{1y}I_{2y}I_{3z}I_{4y},$ $8I_{1y}I_{2y}I_{3y}I_{4z}$, respectively. Then
the effective pure state can be easily constructed from Eq.(13) with the
help of Eq.(14)

$\rho _{eff}=2^{3}D_{0}=\frac{1}{2}E+(2^{3}-1)\times
2^{3}I_{1z}I_{2z}I_{3z}I_{4z}$

$-\stackrel{4}{\stackunder{k=1}{\sum }}\{\exp (-i\frac{\pi }{2}%
I_{kx})WDW\exp (i\frac{\pi }{2}I_{kx})$

$\qquad \times (8I_{1z}I_{2z}I_{3z}I_{4z})\exp (-i\frac{\pi }{2}%
I_{kx})WDW\exp (i\frac{\pi }{2}I_{kx})\}$

$-$ $\stackrel{4}{\stackunder{m>l>k=1}{\sum }}\{\exp (-i\frac{\pi }{4}\times
8I_{kx}I_{lx}I_{mx})WDW\exp [-i\frac{\pi }{2}(I_{kx}+I_{lx}+I_{mx})]$

$\qquad \times (8I_{1z}I_{2z}I_{3z}I_{4z})\exp [i\frac{\pi }{2}%
(I_{kx}+I_{lx}+I_{mx})]WDW\exp (i\frac{\pi }{4}\times 8I_{kx}I_{lx}I_{mx})\}$
\newline
Obviously, to prepare experimentally the effective pure state $2^{3}D_{0}-%
\frac{1}{2}E$ in the four-spin system one needs to perform nine different
experiments directly by starting from the density operator $\sigma
(0)=8I_{1z}I_{2z}I_{3z}I_{4z}$. \newline
\newline
\textbf{(d) A five-spin system}

The density operator $\sigma _{+}(0)$ is chosen as $\sigma
_{+}(0)=2^{4}I_{1y}I_{2z}I_{3z}I_{4z}I_{5z}$,\newline
$2^{4}I_{1z}I_{2y}I_{3z}I_{4z}I_{5z}$,..., $%
2^{4}I_{1z}I_{2z}I_{3z}I_{4z}I_{5y}$; $2^{4}I_{1y}I_{2y}I_{3y}I_{4z}I_{5z}$%
,..., $2^{4}I_{1z}I_{2z}I_{3y}I_{4y}I_{5y}$; \newline
$2^{4}I_{1y}I_{2y}I_{3y}I_{4y}I_{5y}$, respectively. On the basis of
Eqs.(13) and (14) the effective pure state can be readily written as

$\rho _{eff}=2^{4}D_{0}=\frac{1}{2}E+(2^{4}-1)\times
2^{4}I_{1z}I_{2z}I_{3z}I_{4z}I_{5z}$

$-\stackrel{5}{\stackunder{k=1}{\sum }}\{\exp (-i\frac{\pi }{2}%
I_{kx})WDW\exp (i\frac{\pi }{2}I_{kx})$

$\qquad \times (2^{4}I_{1z}I_{2z}I_{3z}I_{4z}I_{5z})\exp (-i\frac{\pi }{2}%
I_{kx})WDW\exp (i\frac{\pi }{2}I_{kx})\}$

$-\stackrel{5}{\stackunder{m>l>k=1}{\sum }}\{\exp (-i\frac{\pi }{4}\times
8I_{kx}I_{lx}I_{mx})WDW\exp [-i\frac{\pi }{2}(I_{kx}+I_{lx}+I_{mx})]$

$\times (2^{4}I_{1z}I_{2z}I_{3z}I_{4z}I_{5z})\exp [i\frac{\pi }{2}%
(I_{kx}+I_{lx}+I_{mx})]WDW\exp (i\frac{\pi }{4}\times 8I_{kx}I_{lx}I_{mx})\}$

$-\exp (-i\frac{\pi }{4}\times 32I_{1x}I_{2x}I_{3x}I_{4x}I_{5x})WDW\exp (i%
\frac{\pi }{2}F_{x})$

$\times (2^{4}I_{1z}I_{2z}I_{3z}I_{4z}I_{5z})\exp (-i\frac{\pi }{2}%
F_{x})WDW\exp (i\frac{\pi }{4}\times 32I_{1x}I_{2x}I_{3x}I_{4x}I_{5x})$ 
\newline
This shows that the effective pure state $2^{4}D_{0}-\frac{1}{2}E$ can be
prepared by performing ($2^{5-1}+1$) different experiments in the five-spin
system. \newline
\newline
\textbf{(e) An arbitrary n-spin (}$\mathbf{n}\geq \mathbf{6}$)\textbf{\
system}

The density operator $\sigma _{+}(0)$ is chosen as $\sigma
_{+}(0)=2^{n-1}I_{1y}I_{2z}...I_{nz}$, \newline
$2^{n-1}I_{1z}I_{2y}I_{3z}...I_{nz}$, ..., $2^{n-1}I_{1z}...I_{n-1z}I_{ny}$; 
$2^{n-1}I_{1y}I_{2y}I_{3y}I_{4z}...I_{nz}$,$...$, \newline
$2^{n-1}I_{1z}...I_{n-3z}I_{n-2y}I_{n-1y}I_{ny}$; $%
2^{n-1}I_{1y}I_{2y}I_{3y}I_{4y}I_{5y}I_{6z}...I_{nz}$,..., \newline
$2^{n-1}I_{1z}...I_{n-5z}I_{n-4y}I_{n-3y}I_{n-2y}I_{n-1y}I_{ny}$; ...... ,
respectively. With the aid of Eqs.(13) and (14) the effective pure state can
be written generally and conveniently as

$\rho _{eff}=2^{n-1}D_{0}=\frac{1}{2}E+(2^{n-1}-1)\times
2^{n-1}I_{1z}I_{2z}...I_{nz}$

$-\stackrel{n}{\stackunder{k=1}{\sum }}\{\exp (-i\frac{\pi }{2}%
I_{kx})WDW\exp (i\frac{\pi }{2}I_{kx})$

$\qquad \times (2^{n-1}I_{1z}I_{2z}...I_{nz})\exp (-i\frac{\pi }{2}%
I_{kx})WDW\exp (i\frac{\pi }{2}I_{kx})\}$

$-\stackrel{n}{\stackunder{m>l>k=1}{\sum }}\{\exp (-i\frac{\pi }{4}\times
2^{3}I_{kx}I_{lx}I_{mx})WDW\exp [-i\frac{\pi }{2}(I_{kx}+I_{lx}+I_{mx})]$

$\times (2^{n-1}I_{1z}I_{2z}...I_{nz})\exp [i\frac{\pi }{2}%
(I_{kx}+I_{lx}+I_{mx})]WDW\exp (i\frac{\pi }{4}\times
2^{3}I_{kx}I_{lx}I_{mx})\}$

$-\stackrel{n}{\stackunder{q>p>m>l>k=1}{\sum }}\{\exp (-i\frac{\pi }{4}%
\times 2^{5}I_{kx}I_{lx}I_{mx}I_{px}I_{qx})WDW$

$\times \exp [i\frac{\pi }{2}%
(I_{kx}+I_{lx}+I_{mx}+I_{px}+I_{qx})](2^{n-1}I_{1z}I_{2z}...I_{nz})$

$\times \exp [-i\frac{\pi }{2}(I_{kx}+I_{lx}+I_{mx}+I_{px}+I_{qx})]WDW\exp (i%
\frac{\pi }{4}\times 2^{5}I_{kx}I_{lx}I_{mx}I_{px}I_{qx})\}$

$-......$\newline
It follows from the above expression that one may need to perform $%
(2^{n-1}+1)$ different experiments by starting from the density operator,
i.e., the longitudinal $n$-spin order component $\sigma
(0)=2^{n-1}I_{1z}I_{2z}...I_{nz}$ in order to prepare experimentally the
effective pure state $2^{n-1}D_{0}-\frac{1}{2}E$ with $n$ qubits. This shows
that the experment number to prepare the n-qubit effective pure state by the
present method is about half number $(2^{n}-1)$ required by the exhaustive
averaging method in Ref.[23]. \newline
\newline
\textbf{3. Discussion }

It has been shown that the effective pure state can be prepared conveniently
on the basis of Eq.(13) when the starting reduced density operator is chosen
suitably. Experimentally, the effective pure state is prepared generally
from the starting density operator, that is, the longitudinal $n$-spin order
component, $\sigma (0)=2^{n-1}I_{1z}I_{2z}...I_{nz}$ in a unified and
systematical form, while the latter can be obtained conveniently from the
equilibrium magnetization in a coupled n-spin (I=1/2) system. Actually, the
starting density operator $\sigma (0)=2^{n-1}I_{1z}I_{2z}...I_{nz}$ can be
obtained easily from the equilibrium magnetization $\sigma _{keq}\propto
I_{kz}$ of any $k$th spin in the coupled spin (I=1/2) system by performing a
sequence of one-qubit quantum gate operations and the two-qubit diagonal
gate operations $G_{kl}(\lambda _{kl})=\exp (-i\lambda _{kl}2I_{kz}I_{lz})$
[26], for example,

$\sigma _{1eq}\propto I_{1z}\dfrac{\pi I_{1y}I_{2z}} {}2I_{1x}I_{2z}\dfrac{-%
\frac{\pi }{2}I_{1y}} {}2I_{1z}I_{2z}\dfrac{\pi I_{2y}I_{3z}} {}\dfrac{-%
\frac{\pi }{2}I_{2y}} {}4I_{1z}I_{2z}I_{3z}\dfrac{\quad \ } {}......$

$\dfrac{\ \quad } {}2^{n-2}I_{1z}I_{2z}...I_{n-1z}\dfrac{\pi I_{n-1y}I_{nz}%
} {}\dfrac{-\frac{\pi }{2}I_{n-1y}} {}2^{n-1}I_{1z}I_{2z}...I_{nz}=\sigma
(0),$ \newline
where the unitary transformation $\exp (-i\theta P)A\exp (i\theta P)=B$ is
denoted briefly as $A\dfrac{\theta P} {}B.$ In the high-temperature
approximation the thermal equilibrium state of an n-spin system can be
written as $\sigma _{eq}\propto \sum_{k}I_{kz}.$ Then $\sigma _{keq}\propto
I_{kz}$ can be obtained by first applying a selective 90$_{x}^{\circ }$
pulse only to the spin $k$, then a hard 90$_{-x}^{\circ }$ pulse to all the
spins of the system, and then applying a magnetic field gradient to the
system to cancel all the transverse magnetizations $I_{ly}$ $(l\neq k)$. The
equilibrium magnetization of any one of the spins in the spin system, e.g.,
the $k$th spin, usually can be enhanced easily by various polarization
transfer processes [11, 24, 25]. Then the present method is a simple scheme
to make use of various polarization transfer techniques to scale up the
qubit size on an NMR quantum computer [11, 21]. Because the equilibrium
magnetization is inverse proportional to the exponential factor $2^{n}$ when
the high-temperature approximation holds in the coupled n-spin system it is
easy to find that the effective pure state prepared by the method has a
signal-to-noise ratio inverse proportional to the factor $2^{n}$ in each
experiment on average, and it is well-known that this is a general
disadvantage of ensemble quantum computation [11, 12, 23]. The present
preparation of the effective pure state involves the unitary operations
including the $n$-qubit Walsh-Hadamard transform $W$, the $n$-qubit
diffusion transform $D$ [5], and the $m$-body $(1\leq m\leq n)$ elementary
propagators exp($\pm i\theta 2^{m-1}I_{k_{1}x}I_{k_{2}x}...I_{k_{m}x})$
built up with the m-body interaction $%
2^{m-1}I_{k_{1}x}I_{k_{2}x}...I_{k_{m}x}$ ($1\leq k_{i}\leq n$, and $%
i=1,2,...,m$) [26, 27]. Each of all these unitary operations is readily
decomposed completely into a sequence of one-qubit quantum gates and the
two-qubit diagonal quantum gates [26]. Therefore, the present method to
prepare the effective pure states could be programmed and performed
conveniently on an NMR quantum computer. \newline
\newline
\textbf{Acknowledgement}

This work was supported by the NSFC general project with grant number
19974064. \newline
\newline
\textbf{References} \newline
1. R.Feynman, Int.J.Theoret.Phys. 21, 467 (1982); Found. Phys. 16, 507
(1986) \newline
2. D.Deutsch, Proc.Roy.Soc.Lond. A 400, 97 (1985) \newline
3. P.W.Shor, \textit{Proceedings of the 35th Annual Symposium on Foundations
of Computer Science}, edited by S.Goldwasser (IEEE Computer Society Press,
Los Alamitos, CA, 1994), p.124; SIAM J.Comput. 26, 1484 (1997) \newline
4. D.Deutsch and R.Jozsa, Proc.Roy.Soc.Lond. A 439, 553 (1992) \newline
5. L.K.Grover, Phys.Rev.Lett. 79, 325 (1997) \newline
6. J.I.Cirac and P.Zoller, Phys.Rev.Lett. 74, 4091 (1995)\newline
7. Q.A.Turchette, C.Hood, W.Lange, H.Mabushi, and H.J.Kimble, \newline
Phys.Rev.Lett. 75, 4710 (1995) \newline
8. D.P.DiVincenzo, Science 270, 255 (1995) \newline
9. Yu.Makhlin, G.Sch$\ddot{o}$n, and A.Shnirman, Nature 398, 305 (1999) 
\newline
10. S.Lloyd, Science 261, 1569 (1993) \newline
11. N.Gershenfeld and I.L.Chuang, Science 275, 350 (1997) \newline
12. D.G.Cory, A.F.Fahmy, and T.F.Havel, Proc.Natl.Acad.Sci.USA 94, 1634
(1997) \newline
13. I.L.Chuang, L.M.K.Vandersypen, X.Zhou, D.W.Leung, and S.Lloyd, Nature
393, 143 (1998)\newline
14. I.L.Chuang, N.A.Gershenfeld, and M.Kubinec, Phys.Rev.Lett. 80, 3408
(1998) \newline
15. J.A.Jones, M.Mosca, and R.H.Hansen, Nature 393, 344 (1998) \newline
16. D.G.Cory, M.D.Price, W.Maas, E.Knill, R.Laflamme, W.H.Zurek, \newline
T.F.Havel, and S.S.Somaroo, Phys.Rev.Lett. 81, 2152 (1998) \newline
17. Z.L.Madi, R.Bruschweiler, and R.R.Ernst, J.Chem.Phys. 109, 10603 (1998) 
\newline
18. N.Linden, H.Barjat, and R.Freeman, Chem.Phys.Lett. 296, 61 (1998) 
\newline
19. M.A.Nielsen, E.Knill, and R.Laflamme, Nature 396, 52 (1998) \newline
20. S.S.Somaroo, C.H.Tseng, T.F.Havel, R.Laflamme, and D.G.Cory, \newline
Phys.Rev.Lett. 82, 5381 (1999)\newline
21. I.L.Chuang, N.Gershenfeld, M.G.Kubinec, and D.W.Leung, Proc.Roy.Soc.
Lond. A 454, 447 (1998) \newline
22. D.G.Cory, M.D.Price, and T.F.Havel, Physica D 120, 82 (1998) \newline
23. E.Knill, I.L.Chuang, and R.Laflamme, Phys.Rev. A 57, 3348 (1998) \newline
24. R.Freeman, \textit{A Handbook of Nuclear Magnetic Resonance} (Longman,
Harlaw, 1987)\newline
25. R.R.Ernst, G.Bodenhausen, and A.Wokaun, Principles of Nuclear Magnetic
Resonance in One and Two Dimensions (Oxford University Press, Oxford, 1987)%
\newline
26. X.Miao, http://xxx.lanl.gov/abs/quant-ph/0003068 (2000) \newline
27. X.Miao, Molec.Phys. 98, 625 (2000) \newline
\newline

\end{document}